\documentclass[12pt]{article}

\def\rdots{\mathinner{\mkern1mu\raise1pt\vbox{\kern1pt\hbox{.}}\mkern2mu
   \raise4pt\hbox{.}\mkern2mu\raise7pt\hbox{.}\mkern1mu}}
\newcommand{\Z}{{\rm Z\kern-.35em Z}}
\newcommand{\bP}{{\rm I\kern-.15em P}}
\newcommand{\Q}{\kern.3em\rule{.07em}{.65em}\kern-.3em{\rm Q}}
\newcommand{\R}{{\rm I\kern-.15em R}}
\newcommand{\h}{{\rm I\kern-.15em H}}
\newcommand{\C}{\kern.3em\rule{.07em}{.65em}\kern-.3em{\rm C}}
\newcommand{\T}{{\rm T\kern-.35em T}}

\newcommand{\be}{\begin{equation}}
\newcommand{\ee}{\end{equation}}

\newcommand{\ve}{\varepsilon}
\newcommand{\la}{\lambda}
\newcommand{\ga}{\gamma}

\newcommand{\pa}{\partial}

\newcommand{\ra}{\rightarrow}

\newcommand{\La}{\Lambda}

\newcommand{\al}{\alpha}

\begin{document}

\openup 1.5\jot
\centerline{For the Quantum Heisenberg Ferromagnet, a Polymer Expansion}
\centerline{and its High \ T \ Convergence}

\vspace{2in}
\centerline{Paul Federbush}
\centerline{Department of Mathematics}
\centerline{University of Michigan}
\centerline{Ann Arbor, MI 48109-1109}
\centerline{(pfed@umich.edu)}

\vfill\eject

\centerline{\underline{Abstract}}
\bigskip

	We let $\Psi_0$ be a wave function for the Quantum Heisenberg ferromagnet sharp in the $\sigma_{zi}$,  and $\Psi_\mu = e^{-\mu H} \Psi_0$.  We study expectations similar to the form

\[   \frac {\left< \Psi_\mu, {\displaystyle\prod_{i \in K}} \sigma_{zi} \Psi_\mu \right>} {< \Psi_\mu, \Psi_\mu >}	\]
for which we present a formal polymer expansion, whose convergence we prove for sufficiently small $\mu$.

The approach of the paper is to relate the wave function, $\Psi_\mu$, to an approximation to it, $\Psi_\mu^{AP}$, a product function
\[	\Psi_\mu^{AP} = \otimes_i \left( \begin{array}{c}
\phi_\mu(i) \\
 \\
1-\phi_\mu(i)
\end{array} \right)_i
\]
where $\phi_\mu(i)$ is a solution of the heat equation on the lattice.  This is shown via a cluster or polymer expansion.

The present work began in a previous paper, primarily a numerical study, and provides a proof of results related to Conjecture 3 of the this previous paper.
\vfill\eject

\section{Introduction.}

This paper continues with some of the concepts presented in a previous primarily numerical study, [1]; but we choose to repeat enough of the definitions to make this paper self-contained.

We consider a lattice, $\La$, and the associated Quantum Heisenberg Hamiltonian
\be	H = - \sum_{i \sim j} (I_{ij} - 1) 	\ee
where $I_{ij}$ interchanges the spins of the two neighboring sites $i$ and $j$ in the lattice $\La$.  We let $p_i$ be the projection onto spin up at site $i$.
\be	p_i = \frac 1 2 \ \big(\sigma_{zi} + 1\big).	\ee
We consider a state $\Psi_0$ with spin up at sites in ${\cal S}_0$, and spin down at the complementary sites.
\be
p_i \; \Psi_0 = \left\{ 
\begin{array}{ll}
\Psi_0 \ , & i \in {\cal S}_0 \\ 
 \\
0 \ , & i \not\in {\cal S}_0 
\end{array}
\right.
\ee
So if  there are $N$ spin ups,
\be	\# \ \{ {\cal S}_0 \} = N .    \ee
We define
\be	\Psi_\mu = e^{-\mu H} \ \Psi_0	\ee
(where we may view $\mu$ as $1/T$, an inverse temperature, or as $t$ an imaginary time).  We introduce an operator, or observable, $A$, as
\be	A = \prod_{i\in K} (p_i + \al)	\ee
for some $\al > 0$; the requirement that $\al \not= 0$ will be clear later.  The objects of study are expectations of $A$.

We set for any observable, $B$
\begin{eqnarray}
[B]_\mu &\equiv& \big< \Psi_\mu, \; B\; \Psi_\mu \big> \\
<B>_\mu &\equiv& [B]/[1] = \frac {\big< \Psi_\mu\;, \;B\; \Psi_\mu \big>}{\big< \Psi_\mu \,, \Psi_\mu \big>}
\end{eqnarray}
Thus we seek to find expressions for $< A >_\mu$, this will be the goal of this paper.

We let $\phi_\mu(i)$ be a solution of the lattice heat equation
\be	\frac \pa {\pa \mu} \ \phi_\mu(i) = ( \Delta \phi_\mu)(i)   \ee
with initial conditions
\be
 \phi_0(i) = \left\{
\begin{array}{ll}
1 \ , & i \in {\cal S}_0 \\ 
 \\
0 \ , & i \not\in {\cal S}_0 
\end{array}
\right.
\ee
We define
\be
\rho_\mu (i) = \frac {(1+\al)\phi^2_\mu(i) + \al(1-\phi_\mu(i))^2}{\phi^2_\mu (i) + (1-\phi_\mu(i))^2}.
\ee
The wave function $\Psi_0$ of equation (1.3) may be written in the explicit representation
\be
\Psi_0 \longleftrightarrow \bigotimes_{i\in {\cal S}_0} \left( \begin{array}{c} 1 \\ 0 \end{array} \right)_i \  
            \bigotimes_{j\not\in {\cal S}_0} \left( \begin{array}{c}0 \\ 1 \end{array} \right)_j
\ee
We introduced in [1] an ``average-field"-like approximate to $\Psi_\mu \; , \; \Psi^{AP}_\mu$ given as.
\be
\Psi^{AP}_\mu \equiv \bigotimes_i \left( 
\begin{array}{c}
\phi_\mu(i) \\
  \\
1-\phi_\mu(i)   \end{array} \right)_i \ .
\ee
The polymer expansion for $<A>_\mu$ will have as a first term
\be
\frac {\left< \Psi^{AP}_\mu, \; A \; \Psi^{AP}_\mu \right>}{\left< \Psi^{AP}_\mu \; , \, \Psi^{AP}_\mu \right>}
\ee
which is easily computed to be
\be	\prod_{i\in K} \rho_\mu(i)	\ee
This is in a special case closely related to the approximation of Conjecture 3 of [1].  The challenge for future research beyond this paper is to get a handle on the remaining conjectures in [1].  It is expected that the constructs of this paper may be sharpened and used to this end; this was the main motivation of the current effort.  We have some ideas towards going beyond the present work.

\vfill\eject

\section{The ``splitting" expansion for $\Psi_\mu$.}
\setcounter{equation}{0}

The expansion of this section is a representation of $\Psi_\mu$ as a sum of product functions (such as the functions of equations (1.12) and (1.13)).  The next section generates the polymer expansion out of the ``splitting" expansion.  We first develop our expansion formally and then deal with its convergence.

We define a {\it nonlinear} mapping from functions on lattice sites to product wavefunctions
\be	{\cal P}(\psi) = \otimes_i \left( \begin{array}{c} 
\psi_\mu(i) \\ 
  \\
1-\psi_\mu(i)   \end{array} \right)_i \ .
\ee
so that referring to equation (1.13)
\be \Psi^{AP}_\mu = {\cal P}(\phi_\mu) \ee
If ${\cal S}$ is a subset of the lattice, we define ${\cal P}_{\cal S}$ to be a similar mapping from functions on the subset ${\cal S}$ of the lattice to product wavefunctions on the associated subset of the Hilbert space.

The ``splitting" expansion is developed from the following computations for two neighboring lattices sites, $i,j$.  We first set
\begin{eqnarray}
d\psi(i) &=& (\psi(j) - \psi(i))d\mu \nonumber \\
 \\
d\psi(j) &=& (\psi(i) - \psi(j))d\mu \nonumber 
\end{eqnarray}
(Differentials here measure changes as $\mu$ increases.)  We then get
\be
(I_{ij} - 1)d\mu {\cal P}_s(\psi) = d{\cal P}_s(\psi) + d\mu \left[ {\cal P}_s(\psi^a) +  {\cal P}_s(\psi^b)- {\cal P}_s(\psi^c) -  {\cal P}_s(\psi^d) \right]
\ee
where $I_{ij}$ is as in equation (1.1) and $s = \{i,j\}$.  We have defined
\begin{eqnarray}
\psi^a &=& \psi \\
\psi^b(i) &=& \psi(j) \\
\psi^b(j) &=& \psi(i) \\
\psi^c(i) &=& \psi^c(j) =  \psi(i) \\
\psi^d(i) &=& \psi^d(j) =  \psi(j)  \ .
\end{eqnarray}
{\it Loosely speaking} the terms in brackets in equation (2.4) tend to cancel in the limit of ``smooth" $\Psi$ (i.e. when $\psi(i) \cong \psi(j) $ ) leading to $(I_{ij} - 1)d\mu {\cal P}_s(\psi) \cong d{\cal P}_s(\psi)$.  In this sense $\Psi^{AP}_\mu$ of equation (1.13) is an exact expression in the limit of ``smooth" $\phi$.

Using equation (2.4) at each edge of the lattice we find the following formal expression.
\be
e^{-\mu H}\Phi_0 = \prod_{s=\{i \sim j\}} \Bigg( \sum^\infty_{n(s)=0} \frac 1{n(s)!} \int^\mu_0 d\mu_{sn(s)} \cdots
\int^\mu_0 d\mu_{s2}  \int^\mu_0 d\mu_{s1}  \sum_{t_{sn(s)}}
\ee
 \[  \cdots \sum_{t_{s2}} \; \sum_{t_{s1}} \Bigg){\rm sign} (\{t\}) {\cal P}(\phi^\bullet_\mu) . \]
Here the initial product is over edges of the lattice.  The integrals over the $\mu_{si}$ in each term in the parentheses may sometimes be looked at as ``time-ordered" with omission of the factorial; this will later be convenient.  The sum over each $t_{si}$ is over the four ``types" of terms in the brackets in equation (2.4), i.e. $t_{si}$ takes on values $a, b, c, d$.  The term ``sign$(\{t\})$" is +1 or -1 depending on whether the total number of $c's$ and $d's$ is even or odd, arising from the signs in equation (2.4).  The ``dot" in $\phi^\bullet_\mu$ represents dependence on ``everything", $\{n(s)\}, \; \{\mu_{sk}\}, \; \{t_{sk}\}$.  The ``component" dependences on $\mu$ imposed in equation (2.3) naturally generalize to $\phi^\bullet_\mu$ satisfying the lattice heat equation.  From equation (2.4) we get the image of a $\phi$ satisfying the heat equation and at some time ``splitting" into four (product) functions, again each satisfying the heat equation till it ``splits".  The generic term in (2.10) is the result of $\displaystyle{\sum_s} n(s)$ ``splittings".  We can now see the equation $\phi^\bullet_\mu$ satisfies, by patiently compiling the result of each ``splitting".

There is the initial condition
\[    \phi^\bullet_0 = \phi_0	\]
and the equation
\be
\frac {\pa}{\pa \mu} \phi^\bullet_\mu = \Delta \phi^\bullet_\mu + \sum_s \sum^{n(s)}_{k=1} \delta(\mu - \mu_{sk}) \cdot Op(s,t_{sk})\phi^\bullet_{\mu^-_{sk}}
\ee
where the $Op$ are operators now defined, and writing $s=\{i \sim j\}$ an (ordered) edge compatible with the definitions used in equations (2.3)-(2.9).

We have set
\[	\phi^\bullet_{\mu^-_{sk}} = \lim_{\mu \rightarrow \mu^-_{sk}} \phi^\bullet_\mu \ .	\]
that is, a limit to $\mu_{sk}$ from below.
\begin{eqnarray}
Op(s,a)\phi &=& 0 \\
(Op(s,b)\phi)(\ell) &=& \Bigg\{
\begin{array}{cc}
0 & \ell \not= i,j \\
\phi(j)-\phi(i) & \ell = i \\
\phi(i)-\phi(j) & \ell = j
\end{array} \\
(Op(s,c)\phi)(\ell) &=& \Bigg\{
\begin{array}{cc}
0 & \ell \not= i,j \\
0 & \ell = i \\
\phi(i)-\phi(j) & \ell = j
\end{array} \\
(Op(s,d)\phi)(\ell) &=& \Bigg\{
\begin{array}{cc}
0 & \ell \not= i,j \\
\phi(j)-\phi(i) & \ell = i \\
0 & \ell = j
\end{array}
\end{eqnarray}

We write $R_\mu$ as the right side of equation (2.10).  By its construction $R_\mu$ ``formally" satisfies
\be
\frac{dR_\mu}{d\mu} = -HR_\mu,  \ \ \ \ R_0 = \Phi_0 \ .
\ee
So likewise ``formally"
\be	\int^\mu_0 (-H R_{\bar\mu}) d\bar\mu = R_\mu - \Phi_0		\ee
But the sums in $R_\mu$ are continuous in $\mu$ and uniformly convergent on compact sets in $\mu$, both viewed in $L^2$, so $R_\mu$ satisfies (2.16) in terms of strong derivatives and so
\be	R_\mu = e^{-\mu H} \ \Phi_0		\ee

In fact since our lattice is assumed finite, equation (2.16) is merely a finite set of coupled linear equations on some $R^n$, constant coefficient no less, so we should expect no difficulty in proving convergence of a formal solution.  But our solution is represented as an infinite sum, and convergence must be checked.  From brief consideration of (2.11) - (2.15) there easily follows:

\noindent
\underline{Basic Theorem}.  Let three sets of initial conditions for the equation (2.11) satisfy
\be    \phi^1_0 \le \phi^2_0 \le \phi^3_0 \ .	\ee
Then for all later $\mu$ the same relation holds
\be    \phi^1_\mu \le \phi^2_\mu \le \phi^3_\mu \ .	\ee
As a special case, if
\be	a \le \phi_0 \le b	\ee
Then
\be	a \le \phi_\mu \le b.	\ee

\vfill\eject

\section{The Polymer Expansion}
\setcounter{equation}{0}

In this section we develop a polymer expansion [2], [3] for the expectation of equation (1.8).  We use the notation and results of [3] (p. 31-38), and we assume the reader has a copy of this before him and is familiar with its nitty-gritty.  The main task is to define the ``polymers", and their ``activity", which will be a technically complex affair.  The decision as to which polymers are ``compatible" will be easy.  The polymer expansions we develop will be formally true for all $\mu$ but convergence will only be proven for $\mu$ small enough.  An extension of the current work to all $\mu$ (so as to address the other conjectures of [1]) would require more than better estimates.  The definitions of the polymers would have to be modified to exhibit cancellation between terms of the expansion using the current polymers.  (In fact we will later use some combinations of the present polymers in estimates in this paper.)  We have some thoughts about polymer definitions needed to go beyond the current paper.  In addition there would be no smallness parameter, here we have $\mu$ for a smallness parameter.  This will make the progress beyond this paper difficult, but we hope to deal with these problems in future work.  We return to the problem at hand.

We will write (using eq. (3.13) of [3]):
\be
\frac{< \Psi_\mu, \Psi_\mu >}{< \Psi^{AP}_\mu, \Psi^{AP}_\mu >} = \sum^\infty_{n=0} \ \frac 1{n!} \  \sum_{\ga_1,...,\ga_n} \prod^n_{i=1} \ z(\ga_i) \; \prod_{j<\ell} \; (1+g(\ga_j,\ga_\ell))
\ee
and similarly:
\be
\frac{< \Psi_\mu, A\Psi_\mu >}{< \Psi^{AP}_\mu, A\Psi^{AP}_\mu >} = \sum^\infty_{n=0} \ \frac 1{n!} \  \sum_{\ga^A_1,...,\ga^A_n} \prod^n_{i=1} \ z(\ga^A_i) \; \prod_{j<\ell} \; (1+g(\ga^A_j,\ga^A_\ell))
\ee
using (some) different polymers in these two expansions (most of the $\ga^A_i$ and $z(\ga^A_i)$ will equal the $\ga_i$ and $z(\ga_i)$).  Then using (3.17) of [3] we will get
\[
\frac{<\Psi_\mu, A\Psi_\mu \>}{<\Psi_\mu, \Psi_\mu>} \Bigg/ \frac{< \Psi^{AP}_\mu, A\Psi^{AP}_\mu >} {< \Psi^{AP}_\mu, \Psi^{AP}_\mu >} = \exp \bigg( \sum^\infty_{k=1} \ \frac 1{k!} \  \sum_{\ga^A_1,...,\ga^A_k} g(\ga^A_1,..., \ga^A_k) \prod^k_{j=1} \ z(\ga^A_j)
\]
\be   - \sum^\infty_{k=1} \ \frac 1{k!} \  \sum_{\ga_1,...,\ga_k} g(\ga_1,..., \ga_k) \prod^k_{j=1} \ z(\ga_j) \bigg) .
\ee
The convergence  of the expression in equation (3.3) will be the subject of the following sections.  One can see that only clusters attached to the variables in $A$ contribute in equation (3.3).

We turn to
\be \frac{< \Psi_\mu, \Psi_\mu >}{< \Psi^{AP}_\mu, \Psi^{AP}_\mu >}   \ee
and substituting the sums from (2.10) for the $\Psi_\mu$ in (3.4), we have a doubled set of sums.  We from here on replace the $\frac 1{n(s)!}$ of (2.10) by a ``time-ordering" of ``vertices".  We label the ``sproutings" in the right sum by $(s,n)$ pairs, and in the left sum by $(\bar s, \bar n)$ pairs.  The inner product is a product of the local inner products at the sites $i$.    We view the left sum in a reversed order so that the $\mu_{\bar s \bar n}$ increase to the right.  Essentially what we will be doing is in $\phi^\bullet_\mu$ separating the contributions of the different ``sproutings".  The synapse-searing aspect of organizing the development is due to the non-linearity of the operation $\cal P$ in equation (2.10).  

\centerline{3.1. \ \ \underline{The Polymers}}

\noindent
\underline{3.1.a}  Comparing to a classical statistical mechanics situation, the {\it particles} of our setup will be the union of {\it left vertices}, $\{(\bar s, \bar n)\}$; {\it right vertices}, $\{(s,n)\}$; and {\it sites}, $\{ i\}$.

\noindent
\underline{3.1.b}  A polymer will contain some number of particles.

\noindent
\underline{3.1.c}   If a polymer contains a right vertex $(s,n)$, it also contains all right vertices at the same edge,  i.e. all  $(s, n')$.  If a polymer contains a left vertex, $(\bar s, \bar n)$, it contains all left vertices at the same edge.  Thus we may say a (left or right) edge belongs to the polymer.

\noindent
\underline{3.1.d}  Two different polymers are {\it compatible} (see [3]) if the particles they contain are disjoint.

\noindent
\underline{3.1.e}  Let $i$ be a site contained in the polymer.  There will be a {\it right $i$-sequence} of length $r(i)$, and a {\it left $i$-sequence} of length $\ell(i)$ associated to $i$.  $r(i)$ and $\ell(i)$ satisfy
\begin{eqnarray}
r(i), \ell(i) &\ge& 0 \\
\max(r(i), \ \ell(i)) &>& 0.
\end{eqnarray}
The right $i$-sequence is of the form
\be	(s_1,n_1), (s_2, n_2), ..., (s_{r(i)}, n_{r(i)})	\ee
and the left $i$-sequence of the form
\be	(\bar s_1, \bar n_1), (\bar s_2,\bar n_2), ..., (\bar s_{\ell(i)}, \bar n_{\ell(i)}) .	\ee
The elements of (3.7) and (3.8) must be particles in the polymer.  The associated $\mu$'s in (3.7) are decreasing and in (3.8) are increasing.  i.e. $\mu_{s_2n_2} < \mu_{s_1 n_1}$, etc.

\noindent
\underline{3.1.f}  Corresponding to (3.7) we introduce $i$-{\it lines} connecting particles as follows
\begin{eqnarray}
\ \ &[i, (s_1,n_1)] & \ \ \nonumber \\
\ \ &	[(s_jn_j), \ (s_{j+1}, \ n_{j+1})] &  \ \ \ \ \ \ j=1,...,r(i) - 1 
\end{eqnarray}
and to (3.8)
\begin{eqnarray}
\ \ &[ (\bar s_{\ell(i)},\bar n_{\ell(i)}), i] & \ \ \nonumber \\
\ \ &	[(\bar s_j \bar n_j), \ (\bar s_{j+1}, \ \bar n_{j+1})] &  \ \ \ \ \ \ j=1,...,\ell(i) - 1 
\end{eqnarray}
The brackets in (3.9) and (3.10) denote lines (unoriented).

\noindent
\underline{3.1.g}  The set of $i$-lines for all $i$ in the polymer must connect the set of its sites and left (right)edges.   The set of particles of the polymer will then be connected by the set of all $i$-lines and automatic connections between left (right)vertices at the same edge.

\noindent
\underline{3.1.h}  A polymer is specified by the particles it contains, and the set of its left and right $i$-sequences (or $i$-lines), of course assuming all conditions above are satisfied.

\underline{Note}:  In a bird's eye view:  the sequence (3.7) represents the contribution to the local inner product at site $i$ of the contributions to $\phi^\bullet$ made at the sprouting $(s_{r(i)}, n_{r(i)})$ as progressing through the other sproutings in the sequence (3.7).

\vfill\eject

\centerline{3.2. \ \ \underline{The Polymer Activity}}
\bigskip

\noindent
\underline{3.2.a} We can view the polymers $\ga$ and $\ga^A$ as being the same (as described in the last subsection) and only their activities being different.  We will write then
\be	z(\ga^A) \equiv z^A(\ga)	\ee
and in this subsection deal with $z$ and $z^A$.

\noindent
\underline{3.2.b}  We will need the lattice Laplacian heat equation Green's function
\be	h(\mu) \cdot \left(e^{\mu\Delta}\right)_{ij}  \equiv g_\mu(i,j)	\ee
where $h$ is the unit step function.

\noindent
\underline{3.2.c}  We define three sets determined by the particle content of the polymer.
\begin{itemize}
\item [$P_L$] = the set of left-edges among the particles.  $\bar s \in P_L$ if left-vertices, $(\bar s, \bar n)$, are particles in the polymer, for all $\bar n$.
\item[$P_R$] = the set of right-edges among the particles.  That is, $s\in P_R$ if right-vertices $(s,n)$ are particles in the polymer, for all $n$.
\item[$P_c$] = the set of sites in the polymer.
\end{itemize}

If $s \in P_R$ we let $n(s)$ be the number of right-vertices at edge $s$.  That is
\be	n(s) = \#\{(s,n)\} 	\ee
where here $s$ is fixed and $n$ may vary but only over right-vertices.  Likewise if $\bar s \in P_L$, then
\be	\bar n(\bar s) = \#\{(\bar s, \bar n) \} \ 	\ee
The definitions of this subsection should be considered referring back to the requirements of subsection 3.1.c.

\noindent
\underline{3.2.d}  We now start defining some objects we will use to specify the activities $\ga$ and $\ga^A$.  We first tabulate values of functions $M_{\mu,i}$ and $M^A_{\mu,i}$, corresponding to $\ga$ and $\ga^A$ respectively.  We will use here notation from 3.1.e.  We first study $M_{\mu,i}$ and its values through a number of cases.

\noindent
\underline{Case 1} \ \ $\ell(i) > 0, \ r(i) > 0$
\be	M_{\mu,i}(x,y) = \frac{xy + (-x)(-y)}{\phi^2_\mu (i) + (1-\phi_\mu(i))^2}	\ee

\noindent
\underline{Case 2} \ \ $\ell(i) = 0, \ r(i) > 0$
\be	M_{\mu,i}(x,y) = \frac{\phi_\mu(i)y + (1-\phi_\mu(i))(-y)}{\phi^2_\mu(i) + (1-\phi_\mu(i))^2}	\ee

\noindent
\underline{Case 3} \  \ $\ell(i) > 0, \ r(i) = 0$
\be	M_{\mu,i}(x,y) = \frac{x\phi_\mu(i) + (-x) (1-\phi_\mu(i))}{\phi^2_\mu(i) + (1-\phi_\mu(i))^2}	\ee
We turn to the definition of $M^A_{\mu,i}$, and refer back to equation (1.6) for the definition of $K$.

\noindent
\underline{Case 1} \ \ $i \not\in K$
\be	M^A_{\mu,i} (x,y) = M_{\mu,i}(x, y)	\ee

\noindent
\underline{Case 2}	\ \ $i \in K, \ \ell(i) > 0, \ r(i) > 0$
\be   M^A_{\mu,i}(x,y) = \frac{(1+\al)xy + \al(-x)(-y)}{(1+\al)\phi^2_\mu(i) + \al(1-\phi_\mu(i))^2} \ee

\noindent
\underline{Case 3}	\ \ $i \in K, \ \ell(i) = 0, \ r(i) > 0$
\be   M^A_{\mu,i}(x,y) = \frac{(1+\al)\phi_\mu (i)y + \al(1-\phi_\mu(i))(-y)}{(1+\al)\phi^2_\mu(i) + \al(1-\phi_\mu(i))^2}
\ee

\noindent
\underline{Case 4}	\ \ $i \in K, \ \ell(i) > 0, \ r(i) = 0$
\be   M^A_{\mu,i}(x,y) = \frac{(1+\al)x\phi_\mu(i) + \al(-x)(1-\phi_\mu(i))}{(1+\al)\phi^2_\mu(i) + \al(1-\phi_\mu(i))^2}
\ee

\noindent
\underline{3.2.e}  We use the notation for $\ga^\bullet$ for $\ga$ or $\ga^A$.  We refer back, particularly to equation (2.10) and 3.2.c, 3.2.d, in the following expression.
\begin{eqnarray}
z(\ga^\bullet) &=& \prod_{\bar s \in P_L} \left( \int^\mu_0 d\mu_{\bar s \bar n(\bar s)} \int^{\mu_{\bar s\bar n(\bar s)}}_0 d\mu_{\bar s (\bar n(\bar s)-1)} \cdot \ \cdot \int^{\mu_{\bar s 2}}_0 d\mu_{\bar s 1} \sum_{t_{\bar s 1}} \cdot \cdot \ \sum_{t_{\bar s \bar n(\bar s)}}\right) \nonumber \\
&\bullet& \prod_{s \in P_R} \left( \int^\mu_0 d\mu_{s1} \int^{\mu_{s1}}_0 d\mu_{s2} \cdot \ \cdot \int^{\mu_{s(n(s)-1)}}_0 d\mu_{sn(s)}  \sum_{t_{sn(s)}}  \cdot \cdot \sum_{t_{ s 1}} \right)  \\
&\bullet& \ {\rm Sign} \ (\{t\}) \ \cdot \ \prod_{i\in P_c} \ M^\bullet_{\mu,i} \big(L(i), R(i)\big) \nonumber
\end{eqnarray}
It remains to define the $L(i)$ and $R(i)$.

\noindent
\underline{3.2.f}  We now define the $R(i)$.  The $L(i)$ are defined in a parallel manner.  

\noindent
\underline{Case 1} \ \ \ $r(i) = 0$.
\be	R(i) = 1	\ee
But in fact the value of $R(i)$ in this case does not matter.

\noindent
\underline{Case 2} \ \ \ $r(i) > 0$.

We use the notation from (3.7) for the right $i$-sequence (an $i$ dependence of the $s$'s and $n$'s there suppressed) and notation from (2.12) - (2.15).  We use hypercompressed notation, explained below.
\be
R(i) = \left( g_{\Delta(1)} Op(1) g_{\Delta(2)} Op(2) \; \cdot \; \cdot \; \cdot \; g_{\Delta(r(i))} Op(r(i)) \phi_{\mu(r(i))}\right)_i	\ee
where we have
\be	g_{\Delta(1)} = g_{(\mu-\mu_{s,n1)}} (\cdot \, , \, \cdot )	\ee
and likewise for the remaining $g's$.

The dots in the arguments of $g$ mean that it is viewed as an operator; equation (3.24) altogether an iterated convolution, leading to a function on the lattice evaluated at site i, as final index outside parenthesis indicates.
\be	Op(1) = Op(s_1, \; t_{s_1n_1})	\ee
and likewise for remaining $Op$'s.
\be	\phi_{\mu(r(i))} = \phi_{\mu_{r(i)n(i)}} .	\ee
\noindent
\underline{Note}:  An important observation that clarifies a number of cogent considerations is the identify
\be	\sum_t \ Op(s,t) = 0 .	\ee
\noindent
\underline{Note}: We have implicitly understood a polymer to be specified by the ``combinatorics" and ``topology" of its connections; the integrals over $\mu$'s and the sum over $t$'s (as in (2.10)) take place for each polymer individually.

\noindent
\underline{Note}:  From the definition of the $M^A$ one can see why $\al$ must be chosen greater than zero.

\vfill\eject

\section{The Superpolymers}
\setcounter{equation}{0}

\centerline{4.1. \ \ \underline{The What}}

We collect polymers with the same values of left edges, $P_L$; right edges, $P_R$; sites, $P_c$; and also the same values of all associated $n(r)$ and $\bar n(\bar r)$.  The sum of such polymers is the {\it superpolymer} associated to $(P_L, P_R, P_c, \{n(r), \bar n(\bar r)\})$.  The activity of a superpolymer is the sum of the activities of the polymers comprising it.  Compatibility conditions are patent.  One may work with these objects instead of the original polymers since they occur as units in expressions (3.2) and (3.3).  The point of dealing with these superpolymers, instead of the original polymers, is that cancellations take place between the polymers in a given superpolymer that are necessary to guarantee convergence.  This will be illustrated by considering a special example in the next subsections.

\centerline{4.2. \ \ \underline{The Why}}

We consider the most special class of polymers with $P_L$ empty, and $P_R$ containing a single edge, $s$.  We let $n(s) = M$, and $\#\{P_c\} = E$.  We further restrict the polymers by limiting ourselves to ones where all $i$-lines (all $E$ of them) contain the first and last vertex (of the $M$ vertices).  For a given value of $i \in P_c$, there are $2^{M-2}$ such possible polymers (counting possible subsets of the $M-2$ remaining vertices).    Then in total there will be $(2^{M-2})^E$ such polymers.  The bound we get on each $R(i)$ is
\be   |R(i)| \le c \cdot g'(i,s)		\ee
where $g'(i,s)$ is a smallness factor extractable if site $i$ is distant from edge $s$ (from the very first $g$ in (3.24).  We will see later we can get
\be	\prod_{i\in P_c} |R(i)| \le \frac c{(E!)^a}	\ee
for some $a$.  This is far from good enough to control the $(2^{M-2})^E$ from the number of terms.

\centerline{4.3. \ \ \underline{The Key}}

The exhibition of the cancelation between terms in a superpolymer is at the heart of the convergence estimates.  We may have been initially disappointed that estimates we could obtain working with our original polymers were not good enough to prove convergence.  But once we discovered superpolymers would work we were more pleased, our results are deeper, more subtle.  Similar to how cluster expansions often converge where perturbation theory does not.  And in fact this will end up being a hard cluster expansion.  Back to the special example at hand.

We wish for a given $i \in R_c$ to consider the sum of $R^\al(i)$ over all polymers described in the previous subsection.  There will then be $2^{M-2}$ terms in the sum.  (Note we are summing over effects in a single $i$-line, keeping the other $i$-lines fixed.)  The $\al$ indicates a labelling of these $2^{M-2}$ polymers.  We then have
\be	| \sum_\al \; R^\al(i)| \le cd'(i,s).	\ee

Understanding bound (4.3) is crucial.  We consider any set of vertices $\cal V$ ordered by their $\mu$ values.  And we look at (for a given $i$ value) all $i$-lines with last vertex, the last vertex in this set, and all vertices elements of this set (i.e. in $\cal V$).  Call the last vertex $v_f$.  Then look at $\displaystyle{\sum_\al} R^\al(i)$, where the summation is over the $i$-lines just described.  Then $\displaystyle{\sum_\al} R^\al(i)$ is the value at site $i$ of the solution at $\mu$ of (2.11) where one sums only over the vertices in $\cal V$ in (2.11) and with their associated $\mu$ values inserted.  The initial condition will be
\be		\phi_{\mu^+_f}  = Op(v_f) \phi_{\mu^-_f}           \ee
$\mu_f$ is the $\mu$ value of vertex $v_f$.  (The $\phi$ on the right side of (4.4) arises from (1.9) - (1.10).) {\it The sum of $R^\al(i)$ in (4.3) is then just the solution of this version of (2.11) by finite iteration.}  And by the Basic Theorem, result (2.2), one has
\be	-  1 \le  \sum_\al R^\al(i) \le 1 .	\ee
The same bounds as on the initial conditions in (4.4).  We urge the reader to work hard to understand this subsection.

\section{Haec Demonstranda Sunt$^*$}
\setcounter{equation}{0}

We take our polymer expansion beyond mere formality . . . this is not string theory . . . by proving convergence (for small enough $\mu$) through the bounds stated in this section, and proved in succeeding sections.  From now on $\ga$ and $\ga^A$ refer to superpolymers.  For a given superpolymer, $\ga$, we define $|\ga|$ to be the number of particles in $\ga$.  The sum we wish to control is
\be
\begin{array}[t]{c}
{\displaystyle\sum_\ga} \\
{\scriptstyle p \in \ga}
\end{array}
|z(\ga)|e^{a|\ga|}
\ee
This is the sum over superpolymers, $\ga$, containing a given particle (vertex or site), $p$.

\noindent
\underline{Bound 5.1}

Let $a$ and $\ve > 0$ be fixed.  Then there is a $\mu_0 = \mu_0(a,\ve)$ such that
\be
\begin{array}[t]{c}
{\displaystyle\sum_\ga} \\
{\scriptstyle p \in \ga}
\end{array}
|z(\ga)|e^{a|\ga|}  \le \mu^{1-\ve}
\ee
if $\mu < \mu_0$.

\noindent
\underline{Bound 5.2}

Let $a, \ve > 0, \ A$, and $\al > 0$ be fixed.  Then there is a $\mu^A_0(a, \ve, A, \al)$ such that
\be
\begin{array}[t]{c}
{\displaystyle\sum_{\ga^A}} \\
{\scriptstyle p \in \ga^A}
\end{array}
|z(\ga^A)|e^{a|\ga^A|}  \le \mu^{1-\ve}
\ee
$\mu < \mu^A_0$.

\vspace{.50in}
\underline{\ \ \ \ \ \ \ \ \ \ \ \ \ \ \ \ \ \ }

$*$ \ \ Passive periphrastic.

\vfill\eject

\section{Some Estimates}
\setcounter{equation}{0}

The bounds of the last section are of a generic type sufficient to prove convergence of a polymer expansion, the ones we intend showing.  We now develop some estimates that will be useful to that end.  Many of these estimates depend on the mechanism of subsection 4.3.

We first turn to the Green's function, $g$, of 3.2.b.  We find it convenient to define
\be	D(x,y) = \max(1, d(x,y))	\ee
where $d(x,y)$ is the distance between sites, or edges, on the lattice.  We present an estimate for the Green's function, on the infinite lattice.

\bigskip
\bigskip
\noindent
\underline{Estimate 6.1}

\begin{eqnarray}
0 \le g_\mu(i,j) &\le& C_N \cdot \min \Big( 1, \mu d^{-N}(i,j) \Big) \\
&\le& \frac{C_N} {D^N(i,j)}
\end{eqnarray}
(for $\mu \le 1)$.  $N$ is an arbitrary integer, and the dependence of $C_N$ on dimension is suppressed.  (We may as well always have this as three.)

\bigskip
\bigskip

Estimate 6.1 arises easily from consideration of the spatial Fourier transform of $g$:
\be	\tilde g = e^{\Delta(k)\mu} \ \delta \ .		\ee
Any derivatives of $\tilde g$ with respect to $k$ bring down at least once factor of $\mu$, and there remains an integral over finite $k$ volume of a bounded function of $k$.

\vfill\eject

\noindent
\underline{Estimate 6.2}

Given $a > 0$, there is an $M=M(a)$, such that
\be		\prod_{i\in \cal S} \frac 1{D^M(i,i_0)} \le \frac{c'_M}{(S!)^a} .	\ee
where
\be	S = \#({\cal S}) .	\ee
\bigskip
\bigskip

Estimate 6.2 is standard, and easy.  One maximizes the left side of (6.5), putting the $S$ $i$'s as close as possible to $i_0$, in a ball centered at $i_0$.  The radius of the ball is $\sim S^{1/3}$, and more than $1/2 \ S$ of the $i$'s are then at a distance from $i_0$ greater than $c S^{1/3}$.  The estimate follows.

\bigskip
\bigskip

For the next few estimates we assume specified a set of vertices $\cal V$.  To a vertex $v$ in $\cal V$ we have associated:

an edge, $s(v)$

an operator, $Op(v)$

of type, $t(v)$

a $\mu$ value, $\mu(v)$

\noindent
We assume all $\mu(v)$ satifsy
\be	\mu > \mu(v) > 0 \ .	\ee
We recall $\phi_\mu$ satisfying (1.9) and (1.10),  and let $\phi^\nu_\mu$ satisfy
\be	\phi^{\cal V}_0 = \phi_0		\ee
and equation (2.11) with the operations from ${\cal  V}$.   We look at $\displaystyle{\sum_\al} R^\al(i)$ where this is sum over all possible $i$-lines with vertices in $\cal V$, but of length $\ge 1$.  (See 3.1.e).  Then
\be	\phi^{\cal V}_\mu(i) = \phi_\mu(i) + \sum_\al R^\al(i)  .    \ee
One has
\be	0 \le \phi_{\mu '} \le 1, \ \ 0 \le \phi^{\cal V}_{\mu '} \le 1, \ \ {\rm all} \ \ \mu' 	\ee
by the Basic Theorem of Section 2.

\bigskip
\bigskip
\noindent
\underline{Estimate 6.3}
\be		-1 \le \sum_\al R^\al(i) \le 1         \ee
and also
\be	|  \sum_\al R^\al(i) | \le \sum_{v \in \cal V} \frac{C'_N}{D^N(i, s(v))} \ .	\ee 
\bigskip
\bigskip

The second estimate here, equation (6.12), is obtained by peeling out the first $g$ in $R^\al(i)$, and using (6.10) to control the remaining sum.  That is we split $\displaystyle{\sum_\al} R^\al(i)$ into a sum over the first operator in each line, and the sum for a given first operator over the rest of the line.  This second sum is controlled by the Subsection 4.3 mechanism (or (6.10)).

\bigskip
\bigskip

Our next esimate we view as central in arguments to follow.  {\it When we first discovered its truth, we ``knew" the polymer expansion could be shown to converge}.  We let $\cal V$ be the set of vertices as in the last estimate.   For a given edge $s$, we will need
\be	n(s) = \# \{ v \in {\cal V} | s(v) = s \} \ .	\ee
We wish to sum over all possible ways $\ell$ different $i$-lines (corresponding to $\ell$ different values of $i$) can be attached to the superpolymer with vertices $\cal V$.  In fact we also will sum over the values of $\ell$.  So we wish to bound
\be	S_T = \sum_\ell \prod^\ell_{r=1} \left( e^f \sum'_{i_r} \left| \sum_{\al_r} R^{\al_r}(i_r) \right| \right) .	\ee
$e^f$ is inserted because we will later want to extract a certain amount of smalless from each $i$-line.  The prime on the intermediate sum indicates all the $i_r$ must be distinct.  We will bound the inner sum by the estimate of equation (6.12).

We will also want to use the standard counting "trick"
\be	|\sum_\beta A_\beta B_\beta|  \le \sum_\beta |A_\beta| \cdot \max_{\beta '} |B_{\beta '}| .	\ee
We write
\be
\left| \sum_{\al_r} R^{\al_r}(i_r) \right| \le \sum_s \frac{C_N \al(s)}{D^4(i_r,s)} \bullet \frac{n(s)}{D^N(i_r,s)\al(s)}
\ee
(with a change in definition of $N$ from (6.12)).  $\al(s)$ will be chosen later.  And
\be	\sum_{i_r} \sum_s \ \frac{C_N \al(s)}{D^4(i_r,s)} \le C \sum_s \al(s) .	\ee
We use here the standard notation of using generic $C$'s that have no important  dependences; $C$ may have different values in different places.  We use (6.15)-(6.17) to get
\be	S_T \le \sum_\ell | C \sum_s \al(s)|^\ell \cdot F(\ell) \cdot \frac 1{\ell !}	\ee	
with
\be	F(\ell) = \max_{\{s_r,i_r\}} \prod^\ell_{r=1} \ \frac{n(s_r)}{\al(s_r)D^N(i_r,s_r)} .   \ee
In (6.18) the $\ell$! is present since each configuration of $\ell$ $i$-lines is counted $\ell !$ times in estimate.  In (6.19) the $i_r$ must be distinct.  We let $|\cal V|$ be the number of vertices in $\cal V$.  One has
\be	|{\cal V} | = \sum_s \ n(s)	. 	\ee

\bigskip
\bigskip
\noindent
\underline{Estimate 6.4}

For $M$ large enough

\be	e^{-M|{\cal V}|} \ S_T \ \le \ C	\ . \ee

Here $M$ may depend on $f$ and no other variables.

\bigskip
\bigskip

We let $w(s)$ be given as
\be	w(s) = \# \left\{ \{s_r, i_r\} | s_r = s \right\} \ee
where the pairs of $\{s_r, i_r \}$ in this expression are those achieving the maximum in (6.19).  One has
\be	\sum_s w(s) = \ell \ .		\ee
We use Estimate 6.2 to show that Estimate 6.4 easily follows from the following.

\bigskip
\bigskip
\noindent
\underline{Estimate 6.5}

There is a $\beta$ and an $M$ such that with $\al_s = (n_s)^{1/(\beta \;+\; 1)}$, and for all $w_s$ and $\ell$
\[	L = e^{-M\Sigma n_s} \; e^{-\ell \; ln\; \ell} \; e^{\Sigma w_s \; ln (\Sigma \al_s)} \cdot  \]
\be	\cdot \  e^{-\beta \Sigma w_s \; ln w_s} e^{\Sigma w_s \; ln \left( \frac{n_s}{ \al_s} \right)} \le C \ .   \ee

\bigskip
\bigskip

We have converted notation $\al(s) = \al_s$, etc.  And of course (6.23) must still hold.

We first use a Lagrange multiplier to maximize with respect to the $w$'s, under the contraint (6.23).  The extremum we are seeking is an interior one.  We also set
\be		\al_s = n^r_s		\ee
and later see why $r = \frac 1{\beta + 1}$ is a nice choice.
\be    ln \; w_s = -1 + \frac 1 \beta \left[ ln(\Sigma n^r_s) + ln \; n^{(1-r)}_s + \la \right] .	\ee
Substituting back into $L$ we find
\be	L \le e^{-M\Sigma n_s} \ e^{-\ell \; ln \ell} \ e^{-(\la - \beta)\ell} .	\ee
We solve (6.26) for $\la$ using (6.23) to get
\be	\la - \beta = \beta \; ln \; \ell - \beta \; ln \; Q		\ee
with
\be	Q = \left(\Sigma n^r_s\right)^{1/\beta} \ \Sigma n_s^{(\frac {1-r}\beta)} \ .	\ee
Setting $r = \frac 1{\beta + 1}$ we now get
\be	Q = \left( \Sigma n_s^{\frac1{\beta+1}} \right)^{\frac{\beta+1}{\beta}}	\ .   \ee
(But other choices of $r$ would also work.)  Therefore
\be
L \le e^{-M\Sigma n_s} \ e^{-\ell \; ln \ell} \ e^{-\beta\ell \; ln \; \ell} \ e^{ \beta \ell \; ln \; Q} .	\ee
Differentiating with respect to $\ell$ to find the maximum:
\be	L \le e^{-M\Sigma n_s} \ e^{(\beta + 1)\bar \ell}	\ee
with
\be	\bar \ell = e^{-1-\frac M{(\beta + 1)}} \left( \Sigma n^{\frac 1{(\beta + 1)}}_s \right) .	\ee
All we need is
\be	M \ge (\beta + 1) e^{-1-\frac M{(\beta + 1)}} \ .	\ee
This proves Estimate 6.5.  There is additional smallness in the $\mu$ factors of the Green's function (see Estimate 6.1) that may be used to control the sum over $\ell$'s in (6.18) to prove Estimate 6.4.  (Perhaps one does not need this extra smallness?)

\bigskip
\bigskip

Our next estimate can be viewed in the setting of summing over possible choices for an $i$-line in ``constructing" a superpolymer;  when there may already be selected a number of $i$-lines, and the line we are adding my connect to one or more disconnected sets of lines already selected.

There are sets of vertices ${\cal V}_1, {\cal V}_2, ..., {\cal V}_n$ and we are summing over $R(i)$ lines where the orderings of ``first appearances" along the line is the order of these sets.  That is, as we move along a given $i$-line if $\mu_r$ is the $\mu$-value for the first appearance of a vertex in ${\cal V}_r$ as one moves along the line, then one has
\be	\mu_{r+1} < \mu_r \ \ \ \ r=1,...,n-1 \ .	\ee
\bigskip
\bigskip
\noindent
\underline{Estimate 6.6}

One then has the bound
\[	|\Sigma R^\al(i)| \le \sum_{u_1,v_1} \sum_{u_2,v_2} \cdots \sum_{u_{n-1}v_{n-1}} \sum_{u_n}
\cdot C^n \cdot    \]
\be  g(i,u_1) g(v_1,u_2) g(v_2,u_3) .... g(v_{n-1}, u_n) 	\ee
where
\be		v_i \in {\cal V}_1 \cup {\cal V}_2 \cup \cdots \cup {\cal V}_i	\ee
\be		u_i \in {\cal V}_i	\ee
and
\be		\mu > \mu(u_1) \ge \mu(v_1) >  \mu(u_2) \ge \mu(v_2) > \mu(u_3) \cdots > \mu(u_n) . \ee

$g(v,v')$ is evaluated at the sites of the vertices and maximized over the four choices (two for each edge).  And more specifically equals $g_{(\mu(v)-\mu(v'))}(\cdot, \cdot).$

We now let $\cal W$ be a set of vertices with $\mu$ values between $\mu$ and $\mu '$, with $\mu ' < \mu$.  We let $Op(s_0, t_0)$ be an operator at $\mu$ value $\mu$.  We look at $i$-line segments beginning with $Op(s_0,t_0)$, with other vertices in $\cal W$, and ending with $Op(s,t)$ at $\mu$ value $\mu_1$, with $\mu_1<\mu '$.  We look at the sum of all such segments
\be	R_s = \sum_\al R^\al_s \ .		\ee
We now consider varying $s$ over all parallel edges on the lattice, but keeping the other parameters the same.  We look at
\be	\sum_s |R_s|		\ee
where here $|R_s|$ is the operator norm of $R_s$ as a mapping from $L^2$ (functions on the lattice at $\mu = \mu_1$) to $L^2$ (functions on the lattice at $\mu = \mu$).

\bigskip
\bigskip
\noindent
\underline{Estimate 6.7}

\be	\sum_s |R_s| \le C	\ee

\bigskip

This estimate will be more useful to us than the previous \underline{Estimate 6.6}, although it may appear more coarse.  It follows with a little thought, from the Basic Theorem, (2.22).

\vfill\eject

\section{The Symmetrization Trick}.
\setcounter{equation}{0}

The reader may view this section as trivial.  But for us it was a crucial epiphany.  We consider summing over all superpolymers containing $N+1$ sites with a distinguished site, $i_0$, fixed through the sum.  Then we may construct this sum by summing
\be	\frac 1 {N!} \ \sum_{\al_0} \ \sum_{i_1} \sum_{\al_1} \sum_{i_2} \sum_{\al_2} \cdots \sum_{i_N} \sum_{\al_N} \ \ ( \ \ \ \ \ \ \ \ )
\ee
where $i_s$ is the site of the $s^{{\rm th}}$ $i$-line, and the $\al_s$ is summed over $i_s$-lines (right or left).  The vertices are introduced as inferred from the $i$-lines.  There are the restrictions that 

a)  the $i$'s are all different.

b)  the $i$-lines are all of length $\ge 1$.

c)  only $i$-line terms in the sum are kept that make the corresponding polymer connected.

\noindent
The ``trick" is  the factor of $1/N!$  in front of equation (7.1).  The price we pay is that as the sums are iteratively done one has to deal with disconnected sub-polymers; e.g. the $i_0$-line and the $i_1$-line may not be connected.  We do not know how to only consider connected objects at all stages, and at the same time employ the Subsection 4.3. mechanism.

\vfill\eject
\section{Counting, Yi Bu Zuo Er Bu Xiu}
\setcounter{equation}{0}

The counting of superpolymer activities to prove the bounds of Section 5 is the final step in verifying convergence of the polymer expansion.  In places we will seem to be sketchy, but the counting is rather technical, if one has been through it many times you know what to worry about.  If it's unfamiliar much chatter will not make it easier.  We make some trivial simplifications
\begin{itemize}
\item[1)] We do not worry about finite volume effects.
\item [2)] We do not worry about the left and right sides of the polymer separately.  Each side separately of a superpolymer may consist of a number of connected pieces, mutually disconnected.  If each of these connected pieces satisfies an estimate similar to the bounds of Section 5, then so does the assembled superpolymer.
\item[3)]  We fix a site $i_0$ for our Section 5 bounds, and do not worry separately about fixing an edge.
\end{itemize}
We count $i$-lines {\it in a particular sequence}, and using Section 7 arguments we can associate $1/N$ smallness to each $i$-line.  (The difference between $N^N$ and $N!$ can be neglected borrowing some smallness at each edge, smallness from $\mu$ factors.  This is not quite trivial, and involves the use of $e^f$ in Estimate 6.4.)

\noindent
\underline{8.1} \ \ We distinguish four types of $i$-lines as introduced in sequence through our counting.

\underline{Type 1}.  This type $i$-line introduces at least one new edge, and is disconnected  to $i$-lines previously introduced.  Note that you may think of the order in which $i$-lines are introduced as opposite to the order in which they are summed.  The second $i$-line is summed over with respect to a fixed first $i$-line, etc.

\underline{Type 2}.  This type $i$-line introduces no new edges, and does not connect any previously disconnected ``pieces" (collections of connected $i$-lines).

\underline{Type 3}.  This type $i$-line may or may not introduce new edges, but it connects  $r \ge 2$ previously disconnected pieces.

\underline{Type 4}.  This type $i$-line introduces some new edges, and is connected  to exactly one extant piece.

\noindent
\underline{8.2} \ \ We here deal with the Type 2 lines introduced.  We suppose there were $\ell$ such lines introduced.  The number of ways of selecting these is
\be	\frac{N!} {\ell !(N-\ell)!}	\ee
The $N$ is as in Section 7.  We have associated a numerical factor
\be	\frac 1 {N^\ell}		\ee
to these $\ell$-lines.  Thus multiplying these we can get $1/\ell !$ smallness to go into Estimate 6.4 with, ending up with a smallness factor after the sum over Type 2 lines, of $\ve^\ell$ from this summation.  Some of the $i$-lines among the $\ell$ of them are not connected to all the vertices in the superpolymer, but to some subset.  This only makes Estimate 6.4 better.

\noindent
\underline{8.3}  \ \ Each time an edge is introduced in the iterative assembly of the superpolymer we introduce all the vertices of the superpolymer at this edge.  We leave over the sum over types and the integral over $\mu$'s.

\noindent
\underline{8.4}  \ \ The decision of what type a line is can be controlled from smallness from the edges.  (Easy but not immediate, similar to in 8.2.)

\noindent
\underline{8.5}  \ \ To take advantage of the mechanism of Subsection 4.3, when a line of each type is introduced so are a number of others that are treated together in estimates (lines of the same root site, lines grouped together start at same site) as follows:
\begin{itemize}
\item[1)] When a type 1 line is summed over so are all type 1 lines with the same edges, introduced in the same order (i.e. the lines all have same arrangement of first appearances).  But see first the note below.
\item[2)] Along with a given type 2 line one introduces all type 2 lines that have their edges in the same extant piece the type 2 line is connected to.  (Each type 2 line is contained (by edges) in one extant connected piece.)
\item[3)]  Each type 3 line is introduced with all type 3 lines connecting the same extant pieces in the same order, and introducing the same new edges, with the same order of first appearances.
\item[4)]  Each type 4 line is introduced at same time as others connected to same extant piece and introducing same new edges with same order of first appearances.
\end{itemize}

\noindent
\underline{Note}.  The groupings we actually use in estimates may correspond to partitioning these groupings into smaller sets, for example when the first appearances occur at specified $\mu$ values.  A particularly egregious example is that we do not need the 4.3 mechanism for type 1 $i$-lines.  The grouping of these terms together is purely for aesthetic reasons.

\bigskip
\noindent
\underline{8.6}  \ \ The counting estimates, particularly those yet to come cost us many sleepless nights.  In the end they were not that difficult.  In the first place, we expected the process to be hard; looking for hard solutions blind you to easy ones.  Secondly, by not seeking any fall-off estimates on the polymers (how the activities depend on the diameter of the superpolymers, for example) we could use Estimate 6.7 and an easy procedure.  To get fall-off behavior seems difficult.  It presumably requires Estimate 6.6 and further bounds of a type we have not considered at all.  The counting is difficult enough though.

\noindent
\underline{8.7}  \ \  Suppose $\bar N$ of the $i$-lines in the construction of the superpolymer are of type 1, 3, or 4.  Let $N_v$  be the number of vertices in the superpolymer.  We can then use
\be	\left( \frac 1{N_v}\right)^{\bar N} \le \left( \frac 1 N \right)^{\bar N} e^{+AN_v}	\ee
for some $A$ (independent of $N_v, \; N, \; \bar N$) to use $\left( \frac 1{N_v}\right)^{\bar N}$ in studying the assembly of the superpolymer through the non-type 2 steps.  (Borrowing some smallness at each vertex to cover the $AN_v$ term.)  By using these factors we can arrange that in the assembly procedure at each stage a connected piece with $n$ vertices in it has associated to it a numerical factor
\be	\frac 1 {N_v} \cdot \frac 1 n	\ee
coming from the factors of $\frac 1 {N_v}$ from (8.3) and some smallness from the vertices in the piece.  We make the following points to help clarify the source of these factors.

a)  A type 1 $i$-line in introduction, if it has $n$ vertices (on the edges connected to it) has a smallness factor $\ve^n$ from which the $1/n$ is borrowed.

\bigskip
\bigskip

\noindent
\underline{Note}.  If the $i$-line is the $i_0$-line then all spatial summations of positions of edges can be done controlled by falloff of $g$ factors.  If it is another $i$-line, one sums over positions of edges but keeps $i$ fixed. When this piece is coordinated with another piece (by a type 3 $i$-line) it is spatially moved as a unit in summing over its $i$-position.  (one uses first the ``position relative to $i$" variables in doing sums when piece arises as a type 1 $i$-line,and sums over $i$ in the step that joins it to other pieces.)

\bigskip
\noindent
\underline{Note}.  All sums in construction of a type 1 $i$-line are easy to control using vertex smallness and $g$ fall-off.

\bigskip

b)	We consider a type 3 $i$-line connecting pieces $\al_1, ..., \al_r$ where piece $\al_i$ has $n_i$ vertices.
To such pieces we have numerical factors associated 
\be	\prod^r_{i=1} \left( \frac 1{N_v} \frac 1 {n_i} \right)    \ee
and the type 3 $i$-line comes through with its own factor of $\frac 1 {N_v}$.  The factors $\left( \frac 1{N_v}\right)^r$ can be traded for a particular choice of $r$ pieces, {\it in a particular order} with a factor of $\frac 1 {\sum^r_{i=1} \; n_i}$ left.  (One can do still better.)  Thus for $r$ pieces selected in a given order we have a numerical factor of $\frac 1{N_v} \cdot \frac 1 {\sum^r_{i=1} \; n_i} \cdot \prod^r_{i=1} \frac 1{n_j}$ left over.   (If there are possibilities $P_\al$ with weights $w_\al$ and $\Sigma w_\al = 1$; we are using $|\Sigma w_\al P_\al| \le \max  P_\al$.  We have had to be only slightly clever to get factor of $\frac 1 {\sum^r_{i=1} \; n_i}$ above.)  We will use the $\Pi \frac 1{n_j}$ factors in summing over possibilities for our type 3 $i$-line, that will connect the $r$ pieces with first appearances in order of the ordering of pieces mentioned.

\bigskip
\noindent
\underline{8.8  \ Example 1}.  We consider the example above in 8.7 b) and assume the type 3 line does not introduce any new edges.  We use the factors of $\Pi \; \frac 1{n_i}$ from above to select the vertex of first appearance for each of the $r$ pieces.  The choice of position for the first first appearance vertex involves a sum
\be 	\sum_s \ g_{(\mu - \mu_1)}(i,s)	\ee
in its estimate.  The remaining sums over positions of first appearance vertices can be bounded using Estimate 6.7.  One is left with numerical factors
\be	C \; C^r \frac 1{N_v} \ \frac 1 {\Sigma n_i}	\ee
very cleanly.

\bigskip
\noindent
\underline{8.9  \ \ Example 2}.  We have dealt with type 2 lines, type 1 lines, and (at least simple) type 3 lines.  Our final example discusses a type 4 $i$-line (and incidentally deals with the complications in a type 3 situation where new edges are introduced).  We use the $1/n$ to select the first appearance vertex in the piece.  We arrange the say $r$ new vertices and the first appearance vertex of the piece in order of decreasing $\mu$ values.  (We only select of the new vertices the $r$ that the $i$-line passes through, there may be others not on the $i$-line.  These can be selected using vertex smallness factors.)  We can sum over positions exactly as in the last example.  This has been a very sleek set of counting estimates, at the price mentioned before:  no spatial fall-off bounds on superpolymer sizes.

\vfill\eject
\section{Results}
\setcounter{equation}{0}

\noindent
\underline{Theorem 9.1}  For each $\ve > 0$ and $A$ (and of course $\al$) one has
\be
\lim_{\mu \ra 0} \frac 1 {\mu^{(1-\ve)} } \left[ \frac{<\Psi_\mu, A\Psi_\mu >}{<\Psi_\mu, \Psi_\mu>} - 
\frac{< \Psi^{AP}_\mu, A\Psi^{AP}_\mu >} {< \Psi^{AP}_\mu, \Psi^{AP}_\mu >} \right] = 0 \ .
\ee
Note expression (1.15) for the second ratio inside the brackets; and that the expression in brackets does not depend on $\al$ if $\#\{K\} = 1$.

This is a generalization of a weakened form of Conjecture 3 of [1].  We do not here pursue the question whether a more careful study of the detailed polymer expressions of this paper would lead to a proof of the exact form of this Conjecture 3.

\bigskip
\bigskip
\noindent
\underline{Acknowledgment}.  I would like to thank Joe Conlon for posing the problem of studying expectations of the type we've been studying, and Elliott Lieb for encouragement.

\vfill\eject

\centerline{\underline{References}}

\begin{itemize}
\item[[1]]  P. Federbush, For the Quantum Heisenberg Ferromagnet, Some Conjectured Approximations, math-ph/0101017.
\item[[2]]  Gallavotti, G., Martin-Lof, A., Miracole-Sole, S.: Some problems connected with the description of coexisting phases at low temperature in the Ising model, Battelle 1971.  In:  Lecture Notes in Physics.  Berlin, Heidelberg, New York: Springer 1971.
\item[[3]] Seiler, E.:  Gauge theories as a problem of constructive quantum field theory and statistical mechanics.
Lecture Notes in Physics.  Berlin, Heidelberg, New York: Springer 1982.

\end{itemize}

\end{document}